%% file: sim_tools_paper.tex
\documentclass[12pt, a4paper]{article}

\usepackage[utf8]{inputenc}
\usepackage[pdftex]{graphicx}
\usepackage{pdfpages}
\usepackage{setspace}
\usepackage[a4paper, left=2.5cm, right=2.5cm, top=2.5cm, bottom=2.5cm]{geometry}
\usepackage[width=0.75\textwidth, labelfont=bf, font=footnotesize]{caption}
\usepackage{url}
\usepackage{hyperref}
\usepackage{nameref}
\usepackage{enumitem}
\usepackage[backend=bibtex, natbib=true, doi=false, isbn=false, url=false, bibencoding=inputenc, bibstyle=chicago-authordate, citestyle=chicago-authordate, maxnames=10, maxcitenames=2]{biblatex}
\begin{document}
	
	\input{./title.tex}
	\onehalfspacing
	\setcounter{page}{2}
	\tableofcontents

\newpage
\section{Introduction}
This paper is a companion paper to the browser-based simulation toolkits of the IS-LM, AD-AS and the Solow growth model. The tools are written by Dominguez-Moran and Geismar (2020) and can be found on GitLab (\url{https://gitlab.tubit.tu-berlin.de/chair-of-macroeconomics}).\footnote{The toolkits were developed as part of a project about innovative teaching. All software is written by Juan Dominguez-Moran (\href{https://twitter.com/osyphys}{@osyphys}) and Rouven Geismar (\href{mailto:r.geismar@outlook.com}{r.geismar@outlook.com}) at the Chair of Macroeconomics at Technische Universität Berlin.} The paper describes how the simulation toolkits can be used to study and teach macroeconomic models and get an intuition for their comparative statics. They facilitate an easy understanding of the underlying economic concepts and the mechanics of the IS-LM, AD-AS and the Solow growth model which are commonly used in academic teaching.
 
Based on the example of the IS-LM model, this paper demonstrates the functionalities and the innovative features of the toolkits.\footnote{For more information about how to use the AD-AS and Solow model toolkits, please see the instruction sections implemented in the respective toolkit.} In addition, the structure of the IS-LM program code, which is distributed under the open source software license GNU AGPLv3, is presented.\footnote{The IS-LM program code and documentation can be found at \url{https://gitlab.tubit.tu-berlin.de/chair-of-macroeconomics}.} The software is implemented using Python and the interactive visualization library bokeh.\footnote{\cite{py} (see also \url{https://www.python.org/}) and \cite{bo} (see also \url{https://docs.bokeh.org/}).}\\

If you use the simulation tools for teaching, we would be grateful for your feedback or a picture of you utilizing the toolkits in class. If you find inspiration in this project for building your own interactive model, we would very much appreciate if you could let us know about your project by mail (\href{mailto:r.geismar@outlook.com}{r.geismar@outlook.com}) or twitter (\href{https://twitter.com/osyphys}{@osyphys}).\\

\section{IS-LM Simulation Toolkit}
\label{ch:2}

\subsection{The basic Model}
The basic version of the IS-LM model explains how output $Y$ and interest rate $i$ are determined in the short-run. Its simplicity and intuititve appeal are reasons for why it is still used in accademic teaching. It captures essential economic phenomena and hence provides a good starting point for teaching economics.
The model describes equilibria on the goods (IS-curve) and the financial (LM-curve) markets. The intersection of both curves determines the short-run equilibrium of the economy. To study the labor market, wage and price developments in the medium- and long-run, one has to use a different model, e.g. AD-AS model.\footnote{A simulation tool for a basic version of the AD-AS model can be found at the Chair of Macroeconomics at Technische Universität Berlin \href{https://www.macroeconomics.tu-berlin.de}{}.} For more details and discussion of the IS-LM framework see \cite{B17} or \cite{BAG17}.

\subsubsection{Goods Market}
Aggregate consumption in the economy is given by $C=A+c(Y-T)$ where $A$ denotes autonomous consumption, $c$ is the marginal propensity to consume, $Y$ denotes aggregate production ($\widehat{=}$ aggregate income) and $T$ are lump-sum taxes. Aggregate investment is defined as $I = B-br$. $B$ denotes autonomous investment, $b$ is the responsiveness of investment to interest rates and $r$ denotes the real interest rate. The Fisher equation is given by $r=i-\pi^e$ where $i$ is the nominal interest rate and $\pi^e$ is expected inflation. The aggregate demand for goods is given by $ZZ=C+I+G+NX$ where $G$ is government spending and $NX$ are net exports. In the short-run equilibrium the demand for goods has to equal production, $ZZ=Y$. Using the previous definitions and rearranging the equation yields the negatively sloped IS-equation

\begin{equation} 
	\mbox{IS-curve:} \quad Y=\frac{1}{1-c}(A+B+G+NX-cT+b\pi^e)-\frac{b}{1-c}i.
\end{equation}

The economic intuition is that an increase in the interest rate leads to a decrease in investment, and hence output. Since output corresponds to aggregate income, people cut back their consumption. This reduction in aggregate demand depresses economic activity further and creates a ‘multiplier effect’. 

\subsubsection{Financial Markets}
The aggregate real money demand is given by $L(Y,i)=h_1Y-h_2i$ where $h_1$ and $h_2$ are the responsiveness of income to money demand and the responsiveness of money demand to interest rates, respectively. The real money supply is denoted by $M/P$ where $P$ denotes the price level. When money supply equals money demand, $M/P=L(Y,i)$, the financial markets are in equilibrium. Because the nominal interest rate is assumed to be greater or equal to zero, the LM-curve exhibits a kink at the zero lower bound:

\begin{equation} 
\mbox{LM-curve:} \quad i=\max\Bigg[0, \enspace{\frac{h_1}{h_2}Y-\frac{M}{h_2P}}\Bigg].
\end{equation}

The intuition for the positive slope of the LM-curve is that an increase in income leads to an increase in money demand (e.g. transaction motive). If the central bank decides to hold the money supply fixed (‘money supply control’) the interest rate has to rise to equilibrate money supply and money demand. If the central bank decides to hold the interest rate fixed (‘interest rate control’) the money supply has to adjust endogenously.

\subsubsection{Goods and Money Market Equilibrium}
An aggregate macroeconomic equilibrium is defined as a state where the goods market and the financial markets are simultaneously in equilibrium. This is the case where the IS and the LM curves intersect.

\subsection{User Interface}
\begin{figure}
	\includegraphics[width=1\textwidth]{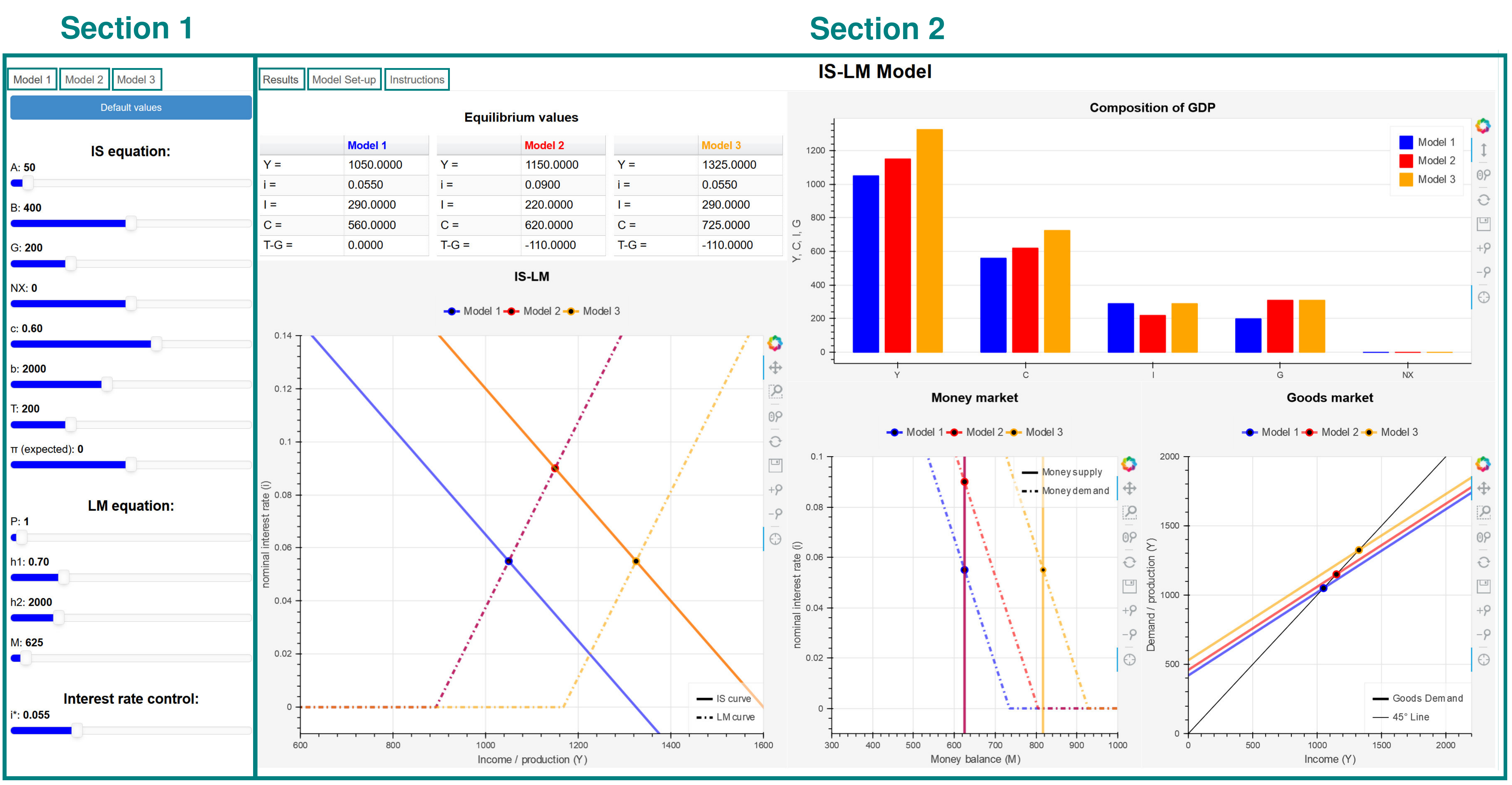}
	\caption{User Interface}
\end{figure}
The user interface (UI) of the program is divided into two major sections. In Section $1$ the model's parameter values (inputs) are chosen. Section $2$ represents the analysis and description section.

\textbf{Section 1:} This section provides the input values for the model. There are three different model tabs (\textit{Model 1} to \textit{Model 3}) which can be used for conducting counterfactual analysis (see \hyperref[sec:ca]{Chapter 2.3}). To distinguish between the Models throughout the user interface, different and unique colors are used for every Model. The blue button on the top restores the default values (\textit{Model 1}) or copies the parameter values of the previous model tab (\textit{Model 2} and \textit{Model 3}). To choose parameter values for the IS or the LM equations the sliders have to be adjusted accordingly. As default, the central bank is assumed to conduct ‘money supply control’, i.e. money supply is set exogenously and the interest rate adjusts endogenously to it. To adjust the interest rate directly, the slider on the bottom labeled ‘interest rate control’ can be used to set the interest rate exogenously. In this case, the money supply adjusts accordingly.

\textbf{Section 2:} This section includes three tabs. The \textit{Results} tab displays all relevant results. It shows the numerical equilibrium values (top-left), the composition of the gross domestic product (top-right) as well as graphical results for the IS-LM Model, the money market and the goods market on the bottom. All graphs are interactive. One can add/remove the different models (\textit{Model} 1 to \textit{Model 3}) by clicking on the respective buttons above the graphs or the bar chart. Different options for manipulating, saving and analyzing the content are placed to the right of each graph. The \textit{Model Set-up} tab shows all relevant model equations and the notation used. The \textit{Instructions} tab provides a brief explanation of how to use the IS-LM-Model program.

\subsection{Counterfactual Analysis - An Example}
\label{sec:ca}
To conduct counterfactual analysis, three models (tabs \textit{Model 1} to \textit{Model 3}) will serve as a ‘playground’. We will now analyze what  happens to the short run equilibrium if the economy experiences an increase in aggregate demand (e.g. an increase in government spending $G$). We will start with the default values given by model 1. In this case, the equilibrium values of the output and interest rate of the economy are given by 1050 currency units (CU) and 5\%, respectively.\footnote{See ‘Equilibrium values’ in Section 2.} Now we use the \textit{Model 2} tab to simulate an exogenous government spending shock. First, we copy the starting values of \textit{Model 1} to \textit{Model 2} by pressing the ‘Assign Values of Model 1’ button on top of Section 1. To see changes to the model happen in real time, we turn on the plots for \textit{Model 2} by pressing the ‘Model 2’ buttons on top of each graph. Now, we increase government spending $G$ to 310 CU by using the slider. This shifts the IS-curve to the right and the deficit increases to -110 CU. If the central bank decides to keep the money supply fixed (‘money supply control’) the interest rate rises to offset the rise in money demand. This leads to crowding out of private investment. Now, suppose the central bank wants to accommodate the increase in the nominal interest rate by increasing the money supply. To implement this policy response, first copy the parameter values of \textit{ Model 2} to \textit{Model 3}.\footnote{Press the button ‘Assign Values of Model 2’ on top of the tab \textit{Model 3}.} Then increase the money supply $M$ by moving the slider as long as the interest rate is back on its starting value of \textit{Model 1}. Alternatively, implement the target interest rate directly by moving the slider ‘Interest Rate Control’. In the latter case, the money supply will adjust endogenously to the interest rate set by the central bank.\footnote{Please keep in mind that the slider ‘Interest Rate Control’ will not update endogenously if money supply changes.} As a result, the LM-curve moves to the right until it crosses the new IS-curve at the old interest rate. This defines the economy's new short-run equilibrium with higher output and private consumption. Now students can compare the effects of the original shock in \textit{Model 2} with the effects of the policy response in \textit{Model 3}.
	
\section{Program Details}
\label{ch:3}

\subsection{Project Structure}

(Quite) Loosely based on the Model-View-Controller software design pattern, the program is composed mainly of:
\begin{itemize}
\item \path{data.py}: model equations and how to calculate initial data
\item \path{input.py}: parameter adjustment widgets coupled to the model update callback in config/interaction.py
\item \path{output.py}: how to plot and show results
\item \path{main.py}: where everything is put together based on arguments specified in \path{config/*.json} and the standalone \path{html/IS_LM.html} is generated
\end{itemize}

The program has no object-orientation. In \path{./demo} there is a version that can run offline since the scripts that are 
usually loaded from the bokeh server have been downloaded and their paths 
specified in the standalone html. The file in \path{config/build_config.py} is an optional tool for creating the config files. Since the whole program makes generalized use of dictionary unpacking when invoking 
functions, it is possible for example to not only change arguments destined for 
plot style, but to add new parameters by looking up their names in the bokeh docs.

\subsection{Requirements}

Prerequisites can be installed using pip as follows:
\begin{verbatim}
pip install -r requirements.txt
\end{verbatim}

\subsection{Running the Program}

The command to create the standalone html at \path{html/IS-LM.html} is
\begin{verbatim}
python3 main.py
\end{verbatim}

\subsection{License}
GNU AFFERO GENERAL PUBLIC LICENSE
Version 3, 19 November 2007\\
Copyright © 2020 Chair of Macroeconomics TU Berlin\\

This interactive IS-LM model is free software: you can redistribute it and/or modify it under the terms of the GNU Affero General Public License as published by the Free Software Foundation, either version 3 of the License or any later version. There should be a copy of the GNU Affero General Public License along with the program. If not, see \url{http://www.gnu.org/licenses/.}

\newpage
\addcontentsline{toc}{section}{References}

\printbibliography

\end{document}

%% file: title.tex
\begin{titlepage}
\begin{center}
Technische Universität Berlin\\
Fakultät VII \\
Institut für Volkswirtschaftslehre \& Wirtschaftsrecht \\
Fachgebiet Makroökonomik \\ [1.00cm]
\includegraphics[width=0.15\textwidth]{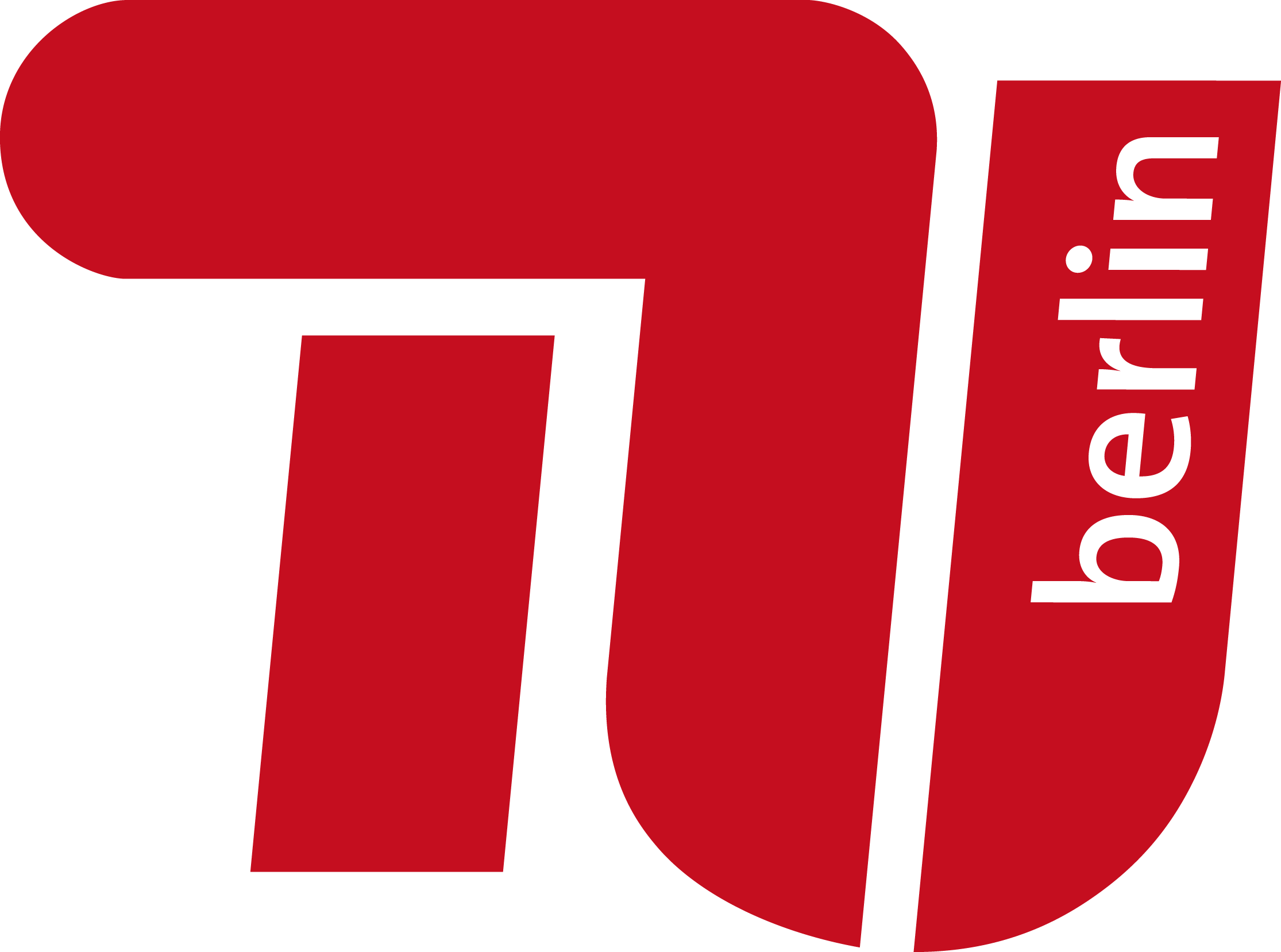} \\[3.0cm]

\textbf{\large 
	Teaching Economics with Interactive Browser-Based Models}\\[0.50cm]

\textit{Companion Paper to the IS-LM, AD-AS and Solow Model Simulation Toolkits } \\[5.0cm]

written by\\[0.80cm]

Juan Dominguez-Moran\\
$<$\href{https://twitter.com/osyphys}{@osyphys}$>$\\[0.25cm]
Rouven Geismar\\
 $<$\href{mailto:r.geismar@outlook.com}{r.geismar@outlook.com}$>$\\[4.50cm]
 
\texttt{1 August 2020} \\

\end{center}
\end{titlepage}